\documentclass[fl eqn,twoside]{article}
\date{}
\usepackage{amsmath}
\usepackage{amsfonts}
\usepackage{amssymb}
\usepackage{graphicx}
\def\be{\begin{equation}}
\def\ee{\end{equation}}

\date{\today}
\begin{document}

\title{{\bf{Equivalence principle in context of large uniform acceleration - a quantum mechanical perspective}}}

\author{
{\bf {\normalsize Sukanta Bhattacharyya}$^{a}
$\thanks{sukanta706@gmail.com}},
{\bf {\normalsize Sunandan Gangopadhyay}
$^{b,c}$\thanks{sunandan.gangopadhyay@gmail.com, sunandan@iiserkol.ac.in, sunandan@associates.iucaa.in}},
{\bf {\normalsize Anirban Saha }
$^{a,c}$\thanks{anirban@associates.iucaa.in}}\\
$^{a}$ {\normalsize Department of Physics, West Bengal State University, Barasat, Kolkata 700126, India}\\
$^{b}$ {\normalsize Department of Physical Sciences, Indian Institute of Science Education $\&$ Research,}\\
{\normalsize Kolkata, Mohanpur 741246, Nadia, India}\\
$^{c}${\normalsize Visiting Associate in Inter University Centre for Astronomy $\&$ Astrophysics,}\\
{\normalsize Pune 411007, India}\\[0.3cm]
}
\date{today}
\maketitle
\begin{abstract}
\noindent 
We study the effect of large acceleration of an uniformly accelerated frame on the validity of weak equivalence principle. Specifically we demonstrate how the behaviour of free quantum particle, as observed by an observer with large uniform acceleration, completely changes from that of a quantum particle emmarsed in a uniform gravitational field. We also extend our analysis to the simplest noncommutative space scenario to show that while spatial noncommutativity does not affect the quantum particle in a gravitational field, it does alter the energy eigenvalues of a quantum particle as seen from a frame with very large uniform acceleration.

%We have studied the quantum mechanics of a free particle in an uniformly accelerated frame. We have constructed the Schr\"{o}dinger equation taking into account the leading order relativistic correction. We then solved this equation exactly to obtain the wave function and the energy eigenvalue. We find that the energy eigenvalue of the particle in an uniformly accelerated frame is identical to that of a $1$-dimensional harmonic oscillator and also admits a zero point energy which owes its origin to the operator ordering. This is in contrast to the result reported earlier in the literature. Here we also investigate the problem in saptial noncommutative field.
\end{abstract}
\vskip 1cm

%%%%%%%%%%%%%%%%%%%%%%%%%\section{Introduction  } %%%%%%%%%%%%%%%%%%%
\section{Introduction}
The physics of uniformly accelerated frame is indistinguishable from that in an uniform gravitational field. This statement forms the core constituent of the equivalence principle (EqP) \cite{paddy} and was originally proposed in the context of classical physics.
%The principle of equivalence \cite{paddy}, that states that the physics of uniformly accelerated frame and that in a uniform gravitational field can not be distinguished, was originally proposed in the context of classical physics.
The validity of this principle was claimed to have been verified in the so called “quantum realm” only much later by Colella, Overhauser and  Werner in a series of experiments known as the Colella-Overhauser-Werner (C--O--W) experiment \cite{COW}-\cite{st} using gravitationally induced quantum-mechanical phase shift in the interference between coherently split and separated neutron de Broglie waves at the 2MW University of Michigan Reactor. The verification was complemented in 1983 by repeating the experiment by Bonse and Wroblewskis in an accelerated interferometer where gravitational effects are compensated \cite{Bonse}. This established that the Schr\"{o}dinger equation in an accelerated frame predicts a phase shift that agrees with observation, as was assumed earlier in the COW experiment \cite{st} for the validity of the EqP in the quantum limit. Since then the EqP in the quantum limit has been verified, time and again, with ever increasing accuracy \cite{litter}. Though, little addressed is the issue of how different the quantum mechanics of a free particle looks when seen from a frame with large uniform acceleration. We shall explore this in this paper.

Also in the last two decade a lot of efforts is spent to construct theories describing high precession experiments in the framework of noncommutative (NC) quantum mechanics (QM) \cite{M. Chaichian, V. P. Nair, B. Chakraborty, P. M. Ho, O. Bertolami, Samanta, A. Saha _2, Stern, A. Saha_3, Anirban Saha, ncgw1, ncgw2} where the standard Heisenberg algebra among the position and momentum operators,
\begin{eqnarray}
[\hat {x}_i,\hat {x}_j]=0, 	~~~~~~~~~~~~~~~~[\hat {p}_i,\hat {p}_j]=0,~~~~~~~~~~~~~~~~~ [\hat {x}_i,\hat{p}_j]=i\hbar \delta_{ij},
\label{20}
\end{eqnarray}
is replaced by the NC Heisenberg algebra
\begin{eqnarray}
[\hat {X}_i,\hat {X}_j]=i\theta_{ij},	~~~~~~~~~~~~~[\hat {P}_i,\hat {P}_j] = 0,~~~~~~~~~~~~~~~~[\hat {X}_i,\hat{P}_j]=i\hbar \delta_{ij},
\label{21}
\end{eqnarray}
with ${\theta_{ij}}$ being an anti-symmetric tensor. NCQM has long been motivated by Gedanken experiments where localizing events in space at Planck scale resolution is considered  \cite{S. Doplicher, D. V. Ahluwalia}.  Although effects of such a NC structure of space may appear near the string/ Planckian scale, it is hoped that some low energy relics of such effects may exist and their phenomenology can be explored at the level of quantum mechanics (QM) \cite{M. Chaichian, A. Saha_3, Anirban Saha, ncgw1, ncgw2, ncgw3, ncgw4, ncgw5}. The C--O--W experiment has been under focus of such NC quantum mechanical modelling in recent years \cite{O. Bertolami, Samanta, A. Saha _2, A. Saha_3, Anirban Saha}. It has been shown in \cite{A. Saha _2, A. Saha_3} that quantizing the Hamiltonian  describing a particle in a constant gravitational field, in a NC space produces a correction term which can be absorbed by a trivial redefination of the momentum operator. Thus spatial noncommutativity does not produce any non-trivial change in the behaviour of a quantum particle in a constant gravitational field. But the complimentary problem of {\it a particle seen from a constantly accelerating frame} has never been approched from NC quantum mechanical platform. We shall investigate this other half of the statement of the equivalence principle, specifically when the acceleration is very large. 

Thus the objective of this paper is twofold. First is to explore to what extent the said EqP in the `quantum limit' holds good when one works with an observer with large uniform acceleration. Second is to see if spatial noncommutativity alters the energy eigenvalues of a quantum particle as seen from a frame with very large uniform acceleration. To achieve this we shall first construct the usual quantum mechanics of a particle seen from an accelerated frame, following which we shall work out the corresponding NCQM in a NC space.

Now to demonstrate the EqP analytically, one can start with the action $-m \int ds$, where $ds$ is line element of a relativistic free particle of mass $m$. In Rindler Coordinates, it is expressed as follows
\begin{eqnarray}
ds^2 = \left (1 + \frac {\alpha x}{c^2} \right)^2 d(ct)^2  -  {dx} ^2  -  {dy}^2  -  {dz}^2 ~ .
\label{1}
\end{eqnarray}
Here $\alpha$ is the uniform acceleration  in the $x$-direction of the Rindler observer.
One can now read off the Lagrangian to be
\begin{eqnarray}
L=-mc^2\left [\left ( 1+\frac{\alpha x}{c^2}\right )^2 -\frac{v^2}{c^2} \right ]^{1/2}~.
\label{2}
\end{eqnarray}
The classical Hamiltonian of this system is therefore
\begin{eqnarray}
H = m c^2 \left  ( 1 + \frac { \alpha x} {c^2} \right ) \left  ( 1 + \frac {\vec{p}^2} {m^2  c^2} \right ) ^{1/2} ~ .
\label{3a}
\end{eqnarray}

Expanding the above Hamiltonian and neglecting the term of the order of $(\frac{1}{c^2})$, i.e. in the low velocity limit, it gives
\begin{eqnarray}
 H = {m} c^2 + \frac {{\vec p}^2} {2m} + m \alpha x .
\label{4}
\end{eqnarray}
The  Hamiltonian (\ref{4}) contains the term $\alpha m x$ which arises due to the uniform acceleration $\alpha$ of the frame and is comparable with the Hamiltonian
\begin{eqnarray}
 H = {m} c^2 + \frac {{\vec p}^2} {2m} +  m g x 
\label{4a1}
\end{eqnarray}
of a particle of mass $m$ in constant gravitational field $\vec{g} = g \hat{i}$. This theoretically confirms the equivalence principle at the classical level under the stated approximation. We can of course quantise both the Hamiltonians (\ref{4}) or (\ref{4a1}) and obtain the Schr\"{o}dinger equation which can either be interpreted as describing the quantum mechanics of a particle from the point of view of an accelerated observer or be interpreted as describing the quantum mechanics of a particle immersed in a constant gravitational field; thus arguing that {\it the equivalence principle works even at the quantum domain.} Note that instead of quantizing Hamiltonian (\ref{4}), we can first quantize the Hamiltonian (\ref{3a}) and subsequently take the low velocity limit to reach the same conclusion. 

However in the expansion of the classical Hamiltonian (\ref{3a}), retaining terms upto the order $(\frac{1}{c^2})$  one obtains
\begin{eqnarray}
 H = {m} c^2 + \frac {\vec p^2} {2m} + m \alpha x + \frac {\alpha x \vec p^2} {2 m c^2}
\label{4a}
\end{eqnarray}
where the fourth term may become important when the acceleration is large and essentially is a deviation from the EqP. This is easily seen at the classical level from the fact that if we recast the Hamiltonian (\ref{4a}) such that it resembles the Hamiltonian (\ref{4a1})
\begin{eqnarray}
 H = {m} c^2 + \frac {\vec p^2} {2m} +  m \tilde{\alpha } x
\label{4b}
\end{eqnarray}
then the redefined effective acceleration $\tilde{\alpha } = \left[ \left( 1 + \frac { \vec p^2} {2 m^2 c^2}\right) \alpha \right]$ becomes momentum dependent. This means two free particles with different momenta will appear to undergo different accelerations from the perspective of a heavily accelerated observer. This is in contrast to the statement of the EqP which says that {\it all free particles will appear to have the same acceleration if seen by an accelerated observer}. Throughout this paper, we want to retain this `next to leading order' term and investigate the ensuing non-trivial effects both in the context of standerd QM and NCQM. 

%%%%%%%%%%%%%%%%%%%%

This paper is organized as follows. In the next section, we will study the quantum mechanical description of a free particle in an accelerated frame in the commutative space. Though it is evident from the discussion above that to the lowest order of approximation, the EqP in the quantum domain is trivially satisfied and the solution to the system in terms of the Airy functions is also well known, we shall see that retaining the `next to leading order term', as has been done in the Hamiltonian (\ref{4a}), will lead to a considerable alteration to the solution. In section \ref{3}, we construct the NCQM of the system and demonstrate the non-trivial effects of spatial noncommutativity. We conclude in section 4.

%%%%%%%%%%%%%%%%%%%%%%%%%%%%%%%%%%%%%%%%%%%%%
\section{Standard Quantum Mechanics of a free particle in an accelerated frame}\label{2a}
%%%%%%%%%%%%%%%%%%%%%%%%%%%%%%%%%%%%%%%%%%%%%

We start by quantizing the Hamiltonian (\ref{4a}) as per the standard prescription which says that the dynamical variables are to be treated as operators that follow the standard Heisenberg algebra (\ref{20}). While it is trivial to follow this dictum for the rest of the terms in the Hamiltonian (\ref{4a}), the next to leading order term poses an ordering ambiguity and a specific choice has to be made.
%Note that, while quantization the Hamiltonian (\ref{4a}) an operator ordering is in order.
Choosing the Weyl ordering prescription, the Schr\"{o}dinger equation for the particle in an uniformly accelerated frame takes the form
\begin{eqnarray}
\hat H \psi = {m} c^2 \psi + \frac {\hat {\vec p} ^2} {2m} \psi + \alpha  m  \hat {x} \psi + \frac {\alpha} {2m c^2} \left [ \frac {1} {3} ( \hat {x} \hat {p}_x^2 + \hat{p}_x^2 \hat{x} + \hat{p}_x  \hat{x} \hat{p}_x) + \hat{x} \hat{p}_y^2 + \hat{x} \hat{p}_z^2 \right] = E \psi.
\label{6}
\end{eqnarray}
This after some algebric manipulation, simplifies to
\begin{eqnarray}
\left [m c^2 + \frac {1} {2m} ( \hat{p}_x^2 + \hat {p}_{y}^2 + \hat {p}_{z}^2) + \alpha  m  \hat{x} + \frac { \alpha} { 2m c^2} (i \hbar \hat {p}_x + \hat{p}_x ^2 \hat{x} + \hat{x} \hat{p}_y ^2 + \hat{x} \hat{p}_z ^2) \right] \psi = E \psi.
\label{7}
\end{eqnarray}
Since the acceleration considered here is only in the $x-$direction, the particle is of course like any other free particle in an inertial frame as far as the $y,z$-directions are concerned. Thus to solve eq.(\ref{7}), we employ the following separation of variables
\begin{eqnarray}
\psi (x,y,z) = N \Phi(x) \exp \left ( -\frac {ip_yy} {\hbar} \right ) \exp \left (-\frac {ip_zz} { \hbar} \right )
\label{8}
\end{eqnarray}
so that eq. (\ref{7}) reduces to a differential equation of unknown function $\Phi(x)$ given by
\begin{eqnarray}
\widetilde {E} \Phi(x) - \frac { \hbar^2} {2m} \frac {d^2 \Phi(x)} {dx^2} - \frac { \hbar^2 \alpha} {2m c^2} \frac{1} {( 1 + \frac { \alpha x} {c^2})} \frac {d\Phi(x)} {dx} = \frac {E\Phi(x)} {( 1 + \frac { \alpha x} {c^2})}~
\label{11}
\end{eqnarray}
where
\begin{eqnarray}
\widetilde{E} = m c^2 + \frac {p_y^2} {2m} + \frac {p_z^2} {2m} ~ .
\label{10}
\end{eqnarray}
Note that since we are looking at a free particle from an accelerated frame with acceleration in the $x$ direction, so $p_y$ and $p_z$ are constants and therefore $\widetilde {E}$  in eq.(\ref{10}) is also a constant. Redefining the independent variable as
\begin{eqnarray}
\xi = 1 + \alpha x/c^2
\label{12}
\end{eqnarray}
eq.(\ref{11}) takes the form
\begin{eqnarray}
\frac {d^2 \Phi(\xi)} {d\xi^2} + \frac{1} {\xi} \frac{d \Phi(\xi)} {d\xi} + \frac { \kappa} {\xi} \Phi(\xi) - \frac { \gamma} {4} \Phi(\xi) = 0.
\label{14}
\end{eqnarray}
Here $ \kappa$ and $\gamma$ are two constants of the system given by
\begin{eqnarray}
\kappa = \frac {2m c^4} { \hbar^2 \alpha^2} E~~,~~ \gamma = \frac {8m c^4} { \hbar^2 \alpha^2} \widetilde{E} ~ .
\label{13}
\end{eqnarray}
Upon a further rescaling of the variable $\xi$ as  $\xi =\gamma^{-1/2}\zeta$, eq.(\ref{14}) takes the form of standard Sturm-Liouville equation
\begin{eqnarray}
\frac{d^2 \Phi(\zeta)} {d\zeta^2} + \frac{1} {\zeta} \frac{d\Phi(\zeta)} {d\zeta} + \left (-\frac{1} {4} + \frac{\sigma}{\zeta}\right )\Phi(\zeta)=0
\label{15}
\end{eqnarray}
where
\begin{eqnarray}
\sigma = \kappa \gamma^{-1/2}
\label{13a}
\end{eqnarray}
is again a constant.
The solution of this equation reads
\begin{eqnarray}
\Phi (\zeta) = \exp \left (-\frac {\zeta} {2} \right )  L_ {\sigma-  \frac {1} {2}} (\zeta)
\label{16}
\end{eqnarray}
where $L_{\sigma-\frac{1}{2}}(\zeta)$ is the Laguerre polynomial.
The wave function (\ref{8}) therefore takes the form
\begin{eqnarray}
\psi ( \zeta,y,z)  = N \exp \left (-\frac { \zeta} {2} \right )  L_{ \sigma - \frac {1} {2}} (\zeta) \exp \left (-\frac {ip_yy} { \hbar} \right ) \exp \left  ( -\frac {ip_zz} { \hbar} \right ).
\label{17}
\end{eqnarray}
Before we go into the discussion of the present solution let us mention that a different solution to this problem of a `free quantum particle in an accelerated frame' has been obtained in \cite{SD_1} where the difference lies in the choice of a different operator ordering.

Coming back to our solution (\ref{17}), we notice that the definition of the Laguerre polynomial demands $(\sigma-1/2)$ to be only a positive integer. This restriction immediately makes the energy of the free particle in an uniformly accelerated frame quantized. Using the expressions for $\widetilde{E}$, $\kappa$, and $\gamma$ from the eqs. (\ref{10}) and (\ref{13}), the quantized energy levels are
\begin{eqnarray}
E_n = \left ( n + \frac {1} {2} \right) \hbar \alpha \frac {\sqrt {p_y^2 + p_z^2 + 2m^2  c^2}} {m c^2}~.
\label{18}
\end{eqnarray}
Given that the motion of the particle in the $y-z$ direction is like any free particle in an inertial frame, it is not surprising that the solution (\ref{17}) is an eigenstate of the operators $\hat{p}_{y}, \hat{p}_{z}$ and will remain so throughout its evolution. Thus we can restrict ourselves to the initial value of $p_y\left(t = 0 \right)=p_z\left(t = 0 \right)=0$ without loosing any non-trivialities. In that case the solution (\ref{17}) will describe the physical situation that the system is moving only along the direction in which Rindler observer is accelerating. Employing the above stated condition, the expression (\ref{18}) yields the quantized energy of the particle to be
\begin{eqnarray}
E_n = \left ( n + \frac {1} {2} \right) \sqrt{2} ~ \hbar \frac { \alpha} {c}  \equiv \left ( n + \frac {1}{2} \right) \hbar \omega
\label{19}
\end{eqnarray}
%The above expression for the energy of the particle in an uniformly accelerated frame is exactly identical to that of a one-dimenssional harmonic oscillator.
%Here the energy eigenvalue given by the eqn. (\ref{19}) shows that if the accelaration $\alpha$ is smaller than the speed of light in vaccum, the contribution in the energy level of the free particle observed from Rindler frame, is noticiablly small to detect.
where $\omega = \frac { \alpha \sqrt {2}} {c}$ is thel frequency. Thus a rather curious feature of the system is that at a considerably large value of acceleration, where the fraction $\frac{\alpha}{c}$ is non-negligible, the energy of the free particle in Rindler frame is not only quantized but also identical, in structure, to that of a 1-D harmonic oscillator.

We would like to elaborate on this to point out the effect of the `next to leading order term' that we have retained in the Hamiltonian (\ref{4a}). Since we have started our calculation from the Rindler metric, which describes the space-time of an accelerated observer, so the analysis that follows is the quantum mechanical description of a free particle from the perspective of the accelerated observer. As long as the acceleration $\alpha$ is small enough, the numerical value $\frac { \alpha \sqrt {2}} {c}$ is so tiny that the discreteness of the quantized energy levels are too small to detect and the energy levels appear to be continuous to the accelerated observer. Within this scenario the `next to leading order term' is negligible and the relevant Hamiltonian is (\ref{4}) and thus the equivalence principle {\it stands}, in the sense that instead of the accelerated observer we can equivalently think of an accelerated particle observed by an inertial observer and both the situation is well-described either by the Hamiltonian (\ref{4}) or by Hamiltonian (\ref{4a1}). Inerestingly the quantum mechanical behaviour in this case completely changes and the solution to this Hamiltonian (\ref{4a1}) comes in terms of the Airy function with energy eigenvalues determined by the roots of the Airy function  \cite{Landau, O. Bertolami, A. Saha _2}.
On the other hand if the acceleration $\alpha$ is very high then the numerical value $\frac { \alpha \sqrt {2}} {c}$ is not negligible, then the energy levels appears discrete with equal spacing $(\Delta E = E_{n+1} - E_{n} = \hbar \omega)$ to the accelerated observer. In this situation the `next to leading order term' is significant and the relevant Hamiltonian is (\ref{4a}) which is not equivalent to (\ref{4a1}) because no matter how high the acceleration $g$ becomes, no `next to leading order term' ever appears in Hamiltonian (\ref{4a1}). So a free particle from the perspective of an accelerated observer is no longer equivalent to an accelerated particle from the perspective of an inertial observer when the acceleration is very high.

Also we observe that there exists a manifest zero point energy in our result which is absent in  \cite{SD_1}. This is not surprising since there a different operator ordering prescription is adopted.

In the next section we will extend our analysis to the noncommutative space and see to what extent the NC effects alter  the results obtained in this section.

%%%%%%%%%%%%%%%%%%%%%%%%%%%%%%%%%%%%%%%%%%%%%%%%%
\section{Quantum Mechanics of a free particle in an accelerated frame in noncommutative space}\label{3}
%%%%%%%%%%%%%%%%%%%%%%%%%%%%%%%%%%%%%%%%%%%%%%%%%
To account for the noncommutative structure of space in our quantum mechanical analysis of the system under consideration, we have to start by quantizing the Hamiltonian (\ref{3a}) using the NC Heisenberg algebra (\ref{21}). Before we proceed further let us note that the operators appearing in the algebra (\ref{20}) and (\ref{21}) are connected by the following set of linear transformations \cite{M. Chaichian}
\begin{eqnarray}
\hat {X}_i=\hat x_i - \frac{1}{2\hbar} \theta \epsilon_{ij}\hat {p}_j~~~~~~~~~~~~~~~~~~,~~\hat {P}_i= \hat{p}_i ~~.
\label{22}
\end{eqnarray}
We will use these relations to express the NCQM in terms of commutative operators which of course follow the standard Heisenberg algebra (\ref{20}). Also to keep the mathematics simple we shall introduce noncommutativity among the direction of the acceleration $\vec{\alpha} = \alpha \hat{i}$ and another spatial direction, say, $\hat{j}$. After Weyl ordering is employed, the resulting Hamiltonian will be 
\begin{eqnarray}
\hat H = [mc^2 + \frac{1}{2m}(\hat{p}_x^2+\hat{p}_{y}^2+\hat{p}_{z}^2)+\alpha m \hat{x}+\frac{\alpha}{2mc^2}(i\hbar\hat{p}_x+
\hat{p}_x ^2\hat{x}+\hat{x}\hat{p}_y ^2+\hat{x}\hat{p}_z ^2) - \frac{\theta \alpha}{2 \hbar c^2}(m c^2\hat p_{y}+\hat p_{y} \hat p^2_{x}+\hat p^3_{y})] ~ .
\label{25}
\end{eqnarray}
Like in the last section, here, while taking the approximation we have retained terms upto order $\frac{1}{c^2}$. The last term  of the left hand side in (\ref{25}) proportional to $\theta$ represents the effect of spatial noncommutativity. Using the same seperation of variables % separable form
 for the wave function as in the previous section
\begin{eqnarray}
\psi(x,y,z) =N \Phi(x) \exp\left (-\frac{ip_yy}{\hbar}\right )
\exp\left (-\frac{ip_zz}{\hbar}\right ) 
\label{26}
\end{eqnarray}
the time-independent Schr\"{o}dinger equation for the Hamiltonian (\ref{25}) reads
\begin{eqnarray}
\hat{H}_{NC} \phi(x)  = (\hat{H}+ \hat{H}^\prime)\Phi(x)=E \Phi(x)
\label{28}
\end{eqnarray}
where 
\begin{eqnarray}
\hat{H}  \Phi(x) = (mc^2 + \frac{p_y^2} {2 m} + \frac{p_z^2} {2 m}) (1 + \frac{\alpha x} {c^2}) \Phi(x) - \frac{\hbar^2} {2 m} (1 + \frac{\alpha x} {c^2}) \frac{d^2\Phi(x)} {dx^2} - \frac{\hbar^2 \alpha}{2 m c^2} \frac{d\Phi(x)}{dx}
\label{unptb_H}
\end{eqnarray}
represents the unperturbated system and 
\begin{eqnarray}
\hat{H}^\prime \Phi(x) = \frac{\alpha \theta m} {2 \hbar}p_{y} \Phi(x) + \frac { \alpha \theta p^3_{y}} {4 m \hbar c^2} \Phi(x) - \frac{\alpha \theta \hbar} {2 c^2} p_{y} \frac{d^2\Phi(x)}{dx^2}
\label{29}
\end{eqnarray}
describes the perturbation which arises due to introduction of noncommutativity among the $x-y$ coordinates.

 Now we calculate the change in the ground state energy due to the time independent perturbation which arises from the effect of spatial noncommutativity. We already have the unperturbed wave function from eqn. (\ref{16}) of the previous section which, for the groud state is given by
\begin{eqnarray}
\Phi_{0}(\zeta)=\exp(-\frac{\zeta}{2})
\label{30}
\end{eqnarray}
since the  Laguerre polynomia $L_{0}(\zeta)=1$. To first order in the perturbation, the shift of the ground state energy eigenvalue is given by
\begin{eqnarray}
E^\prime_0 = \frac{\alpha \theta m}{2 \hbar}(1 + \frac{p^2_y}{2 m^2 c^2}) p_y ~ .
\label{31}
\end{eqnarray}
We thus obtain a nontrivial shift in energy of the particle owing to noncommutativity. It is to be noted that not only the energy shift in the ground state observed by an accelerated frame is momentum dependent, moreover, it is the non-zero $p_{y}$ that causes this shift to appear. This is not surprising because if all aspects of the motion concerned, e.g. the momentum and acceleration are confined to the $x-$direction then the particle is effectively in one dimension and thus spatial noncommutativity can not affect it. 

We further observe that it is easy to show that the effect of the spatial noncommutativity in the Hamiltonian (\ref{4a1}) can be readily absorbed by a momentum re-definition, making such noncommutative effect trivial . Thus quantum mechanical description of a particle under constant gravitationl field $\vec{g} = g\hat{i}$ does not get modified due to spatial noncommutativity. By the same token, the Hamiltonian (\ref{4}) for the free particle from the perspective of the accelerated observer, in its lowest order approximation does not get modified due to spatial noncommutativity. However, once we include the `next to leading order term' in the Hamiltonian (that is the term of ${\cal{O}}(\frac{1}{c^2})$), the spatial noncommutativity begins to play a non-trivial role as the effect of noncommutativity can no longer be absorbed by a momentum rescaling in the Hamiltonian (\ref{4a})).

\section{Concluding Remarks} \label{3a_1}
In this paper, we have analysed the quantum mechanics of a free particle observed from a frame with large uniform acceleration both in commutative and noncommutative space. For this purpose, first we started with the line element of a relativistic free particle and expressed it in Rindler coordinates. From the line element in the Rindler coordinates, we read off the Lagrangian and obtained the corresponding Hamiltonian. Expanding this Hamiltonian in the low velocity limit i.e., neglecting the terms of the order of $(\frac{1}{c^2})$ and higher, one can demonstrate the weak equivalence principle which says that a free particle from the perspective of an accelerated observer is equivalent to a particle emmersed in a constant gravitational field.

To examine the deviation from the equivalence principle when the acceleration of the observer is large, we instead, retain terms upto the `next to leading order' while expanding this Hamiltonian and investigate its effect both in the context of standard quantum mechanics and noncommutative quantum mechanics. That the Hamiltonian, thus obtained, shows a violation of the equivalence principle can be readily observed by noting that the effective acceleration becomes momentum dependent, already at the classical level.

 To proceed to the quantum mechanical analysis of this problem, the dynamical variables are treated as operators and the Weyl ordering prescription is taken into account. After solving the Schr\"{o}dinger equation in the commutative space, we get the wave functions in terms of the Laguerre polynomial. We observed that the energy of the system is not only quantized but is also identical to that of a 1-D harmonic oscillator ones we make the system confined to one dimension by choosing the appropriate initial conditions, namely,  $p_y\left(t = 0 \right)=p_z\left(t = 0 \right)=0$. Thus, if the acceleration of the observer is considerably large, then the `next to leading order term' is significant and the relevant Hamiltonian is not equivalent to that of a particle in a constant gravitational field. This is because no matter how high the gravitational acceleration $g$ becomes, no `next to leading order term' ever appears in the corresponding Hamiltonian. So a free particle from the perspective of an observer with large uniform  acceleration is no longer equivalent to a particle with constant acceleration (due to gravity) from the perspective of an inertial observer. The existence of the zero point energy of the system is also notable and owes its origin to the Weyl ordering prescription adopted in this work. Our analysis differs from \cite{SD_1} in this regard. 
 
 Extending our analysis in noncommutative space we find that there exists a non-trivial momentum-dependent shift in the energy spectrum of the free particle observed from an uniform accelerated frame. This is in contrast to the case of a particle under constant gravitational field in noncommutative space where one finds no change in the energy spectrum.

%%%%%%%%%%%%%%%%%%%%%%%%%%%%%%%%%%%%%%%%%%%%%%%%%%%%%%%%%%%%%%%%%%

\section*{Acknowledement}
S.G. acknowledges the support by DST SERB under Start Up Research Grant (Young Scientist), File No.YSS/2014/000180. AS and SB acknowledges the financial support of DST SERB under Grant No.SR/FTP/PS-208/2012.

%%%%%%%%%%%%%%%%%%%%%%%%%%%%%%%%%%%%%%%%%%%%%%%%%%%%%%%%%%%%%%%%%%

\end{document}